\documentclass[]{aastex631}
\usepackage{xcolor}
\usepackage{amsmath}
\usepackage{derivative}
\let\tablenum\relax 
\usepackage[print-unity-mantissa=false]{siunitx}
\usepackage{bm}
\usepackage[safe]{tipa}

\newcommand{\okina}{\textquoteleft}

\newcommand{\FeH}{[\text{Fe}/\text{H}]}

\newcommand{\alphaFe}{[\alpha/\text{Fe}]}
\newcommand{\fHHO}{f_{\mathrm{H}_{2} \mathrm{O}}}

\DeclareSIUnit\parsec{pc} 
\DeclareSIUnit\year{yr}

\received{2023 July 20}
\revised{2023 October 13}
\accepted{2023 October 14}
\published{2023 November 14}
\reportnum{AJ 166 241}

\shorttitle{The Galactic ISO Population}

\shortauthors{Hopkins et al.}

\graphicspath{{./}}

\begin{document}

\title{The Galactic Interstellar Object Population: A Framework for Prediction and Inference}

\correspondingauthor{Matthew Hopkins}
\email{matthew.hopkins@physics.ox.ac.uk}

\author[0000-0001-6314-873X]{Matthew J. Hopkins}
\affiliation{Department of Physics, University of Oxford, Denys Wilkinson Building, Keble Road, Oxford OX1 3RH, UK}

\author[0000-0001-5578-359X]{Chris Lintott}
\affiliation{Department of Physics, University of Oxford, Denys Wilkinson Building, Keble Road, Oxford OX1 3RH, UK}

\author[0000-0003-3257-4490]{Michele T. Bannister}
\affiliation{School of Physical and Chemical Sciences|Te Kura Mat\={u}, University of Canterbury,
Private Bag 4800, Christchurch 8140,
Aotearoa New Zealand}

\author[0000-0001-8108-0935]{J. Ted Mackereth}
\altaffiliation{Banting Fellow}
\affiliation{Just Group plc, Enterprise House, Bancroft road, Reigate, Surrey RH2 7RP, UK}
\affiliation{Canadian Institute for Theoretical Astrophysics, University of Toronto, 60 St. George Street, Toronto, ON M5S 3H8, Canada}
\affiliation{Dunlap Institute for Astronomy and Astrophysics, University of Toronto, 50 St. George Street, Toronto, ON M5S 3H4, Canada}

\author[0000-0002-1975-4449]{John C. Forbes}
\affiliation{School of Physical and Chemical Sciences|Te Kura Mat\={u}, University of Canterbury,
Private Bag 4800, Christchurch 8140,
New Zealand}

\begin{abstract}
The Milky Way is thought to host a huge population of interstellar objects (ISOs), numbering approximately \(\qty{1e15}{\per\parsec\cubed}\) around the Sun, which are formed and shaped by a diverse set of processes ranging from planet formation to Galactic dynamics. 
We define a novel framework: firstly to predict the properties of this Galactic ISO population by combining models of processes across planetary and Galactic scales, and secondly to make inferences about the processes modelled, by comparing the predicted population to what is observed.
We predict the spatial and compositional distribution of the Galaxy's population of ISOs by modelling the Galactic stellar population with data from the APOGEE survey and combining this with a protoplanetary disk chemistry model.
Selecting ISO water mass fraction as an example observable quantity, we evaluate its distribution both at the position of the Sun and averaged over the Galactic disk; our prediction for the Solar neighbourhood is compatible with the inferred water mass fraction of 2I/Borisov.
We show that the well-studied Galactic stellar metallicity gradient has a corresponding ISO compositional gradient. 
We also demonstrate the inference part of the framework by using the current observed ISO composition distribution to constrain the parent star metallicity dependence of the ISO production rate.
This constraint, and other inferences made with this framework, will improve dramatically as the Vera C. Rubin Observatory Legacy Survey of Space and Time (LSST) progresses and more ISOs are observed.
Finally, we explore generalisations of this framework to other Galactic populations, such as that of exoplanets. 
\end{abstract}


\section{Introduction} 
\label{sec:intro}

1I/\okina Oumuamua \citep{Meech_2017} and 2I/Borisov\footnote{\url{https://minorplanetcenter.net/mpec/K19/K19RA6.html} and \url{https://minorplanetcenter.net/mpec/K19/K19S72.html}} are the first two observed samples from a highly numerous population: interstellar objects (ISOs). 
Estimated to number \(\sim \qty{1e15}{\per\parsec\cubed}\) around the Sun \citep{Engelhardt_2017,Do_2018}, they are implied to have a spatial distribution spanning the entire Galaxy. 
This population has been predicted to exist for decades \citep{McGlynn_1989}, based on models of the accretion and migration of the giant planets, which predict that 75–85\% of cometary bodies initially in the Solar System must have been scattered into interstellar space \citep{Fernandez_1984,Brasser_2006}. 
Modern exoplanet surveys consistently find that giant planets are common across the Galaxy around stars with a range of spectral types \citep{Fulton_2021, Sabotta_2021}.
This makes planetesimal scattering common across the Galaxy. 
A significant number of planetesimals can also be ejected by close stellar flybys early in a planetary system's life \citep[e.g.][]{Pfalzner_2021}. 
The protoplanetary disks of other stars are therefore expected to be a source of ISOs \citep{Stern_1990, Moro-Martin_2022}.

Initially it was expected that interstellar objects would display cometary characteristics \citep[e.g.][]{Jewitt_2003}. 
The population's dominant dynamical formation mechanisms would preferentially harvest more distant, ice-rich planetesimals from the disks of the source systems.
More of the cometary ISO population passing through the Solar System could be detected than rocky ISOs, as comae  make cometary ISOs brighter down to smaller diameters \citep{Engelhardt_2017}.
2I/Borisov appeared relatively similar in size and composition to Solar System comets \citep{Jewitt_2023}. 
Its distinctive features were an exceptionally high CO and NH$_2$ content, implying it formed on the edge of its home system's protoplanetary disk, beyond the hypervolatile CO ice line \citep{Bodewits_2020, Cordiner_2020}. 

However, 1I/\okina Oumuamua had a mix of observed characteristics.
A 160-m scale object, it lacked a coma in deep imaging \citep{Jewitt_2017}  or detectable outgassing in CO, C\(\mathrm{O}_2\) \citep{Trilling_2018} or CN \citep{Ye_2017}. 
Despite this, it still underwent non-gravitational acceleration similar to that experienced by Solar System comets \citep{Micheli_2018}. 
The large amplitude of its light curve implied a high-aspect-ratio shape \citep{Mashchenko_2019}. 
At the time, this combination of factors seemed unusual, although the surface reflectance properties were consistent with outer Solar System bodies \citep[e.g.][]{Bannister_2017}.
Hypotheses for the composition and formation of 1I/\okina Oumuamua that match the limited data remain varied. 
It may be a planetesimal \citep{'OumuamuaISSITeam_2019};
a fragment of a comet devolatized by passages close to its parent star before ejection \citep{Raymond_2018b};
an icy fractal aggregate \citep{Moro-Martin_2019}; 
a hydrogen iceberg formed in a molecular cloud \citep{Seligman_2020};
or a nitrogen ice fragment from the surface of a Pluto-like dwarf planet  \citep{Jackson_2021}. 
Recently, observation of similarly-small near-Earth asteroids have identified objects with the lightcurve amplitudes seen in 1I.
Additionally, \cite{Farnocchia_2023} and \cite{Seligman_2023} report six asteroids with significant non-gravitational acceleration and no coma.
While from the two ISOs found so far, the compositions of interstellar objects are clearly varied, 1I may be less extreme than it was first considered.

ISOs formed in a protoplanetary disk carry information about their home systems in their composition: we have samples of other planetary systems coming to us for study. 
The composition of a protoplanetary disk correlates with the elemental abundances of its central star, due to their formation from the same gas and dust in a molecular cloud core \citep{Oberg_2021}. 
We can thus expect stars of different metallicities to produce ISOs of different compositions.
The composition of this gas and dust varies in both space and time as the Galaxy chemically evolves due to stellar nucleosynthesis, making the current Galactic ISO population sensitive to the entire history and evolution of the Milky Way over cosmic time \citep{Tinsley_1974, Lintott_2022}. 
Since the occurrence of planetesimal-scattering giant planets also has a metallicity dependence \citep{Fischer_2005}, the relative occurrence of ISOs of different compositions will therefore carry information about the Galaxy's distribution of planetary architectures. 
Finally, we expect many ISOs to outlive their parent stars, as they occupy a similar environment to Gyr-old Oort cloud comets.
At minor-planet sizes, ISOs are not subject to any known destructive forces in the ISM \citep{Guilbert-Lepoutre_2015}, other than minor erosion by dust, in their frequent passages through molecular clouds \citep{Pfalzner_2020}.
They could also be disrupted or devolatilised in rare close encounters with stars \citep{Raymond_2018b, Forbes_2019}. 
ISOs thus present the possibility of studying long-lost planetary systems.

The Vera C. Rubin Observatory Legacy Survey of Space and Time (LSST) \citep{Ivezic_2019} is predicted to provide a sample of tens of 1I/\okina Oumuamua-like ISOs \citep[e.g.][]{Levine_2021}, as well as any more 2I/Borisov-like cometary ISOs that enter our Solar System. This is in addition to the continuing contributions of the NEO surveys and other observatories that found 1I and 2I in the first place \citep{Meech_2017}. 
These will provide a large and varied sample of interstellar objects, for study and comparison to predictions.

The many dependencies over planetary and Galactic scales make the observed ISO population a fascinating tool: it has the potential to test models in both Galaxy and planetary physics, with an entirely different set of biases to traditional methods. 
\cite{Lintott_2022} introduced the concept of predicting the composition of a galaxy's population of ISOs from the galaxy's stellar distribution. 
Using simulated galaxies from EAGLE \citep{Schaye_2015}, they showed that the water content of ISOs was sensitive to galactic star formation history. 
In this work, we develop this method and apply it to the stellar population of the Milky Way, estimated with data from the APOGEE survey, to predict a broader set of properties of our own Galaxy's population of interstellar objects. 
We predict the distribution of ISOs in both their spatial position in the Galaxy and their water mass fraction. 
By evaluating this distribution at the current position of the Solar System, we predict the properties of the population of ISOs from which the observed sample of the 2020s will be drawn, and we compare this to the whole-Galaxy distribution. 
We then detail a Bayesian method of comparing the predicted and observed distributions to make inferences about the planetary and Galactic processes modelled, and demonstrate this method by constraining the metallicity dependence of the ISO production rate.

\section{APOGEE and Stellar Density Modelling}\label{sec:data}
To predict the distribution of ISOs in the Milky Way, we first obtain the distribution of all stars throughout Galactic history over a large swath of the Galactic disk, which we model by fitting simple density profiles to debiased data from the APOGEE survey.
While APOGEE's main sample is not representative of all extant stars, we can extrapolate from it to recover the total stellar population of the Milky Way.
By design, the APOGEE main sample mainly contains red giants: stars in a relatively short-lived phase towards the end of their lives. 
Since each stellar generation forms stars with a range of masses \citep{Chabrier_2003} and a corresponding range of lifespans, all but the newest stellar populations will have some stars currently in the red giant stage.
This means multiple generations are represented in the APOGEE sample.
We then extrapolate to the entire stellar population.
This reconstruction is detailed in \ref{sec:sinemorte}.

\subsection{Observational Data: APOGEE}
We use data from the APOGEE SDSS-IV Data Release 16 \citep{Jonsson_2020}. 
APOGEE is a near-infrared, high-resolution (\(R\sim \num{22500}\)) spectroscopic stellar survey used to estimate high-precision chemical abundances for a sample of over \(\num{200000}\) Milky Way stars \citep{Majewski_2017}.
The survey's simple and well-characterised selection function, based on apparent magnitude and dereddened colour, is optimised to select red giants \citep{Zasowski_2013, Zasowski_2017}. 
As infrared bands suffer less from dust extinction, the red giants selected are visible across the Galactic disk. 
The high-precision abundance measurements mean that we can identify monoabundance populations with very low levels of contamination by binning the APOGEE stars in \(\FeH\) and \(\alphaFe\) \citep{Bovy_2016a}.
Additionally, the well-characterised nature of the selection function means that it can easily be accounted for using the method of \cite{Bovy_2016b}.
This makes APOGEE an ideal choice for modelling the spatial and chemical distribution of the Milky Way's red giant population.

We obtain the chemical abundances, heliocentric distance and age of stars in APOGEE DR16.
To obtain each star's abundances, we use the calibrated ASPCAP pipeline's \citep{GarciaPerez_2016} abundances of iron \(\FeH\) and alpha elements \(\alphaFe\), calculated from an average of the abundances of the elements O, Mg, Si, S, and Ca, after \cite{Bovy_2016a}. 
For each star's heliocentric distance and age, we use the \texttt{weighted\textunderscore dist} and \texttt{age\textunderscore lowess\textunderscore correct} estimate of \texttt{AstroNN} \citep{Leung_2019, Mackereth_2019}. 
\texttt{weighted\textunderscore dist}  is a weighted average of the distance estimate from the \textit{Gaia} parallax measurement of the star \citep{GaiaCollaboration_2016} and a spectro-photometric distance estimate of the star from a neural network trained on \(\num{265761}\) stars surveyed in common between APOGEE DR14 \citep{Abolfathi_2018, Holtzman_2018} and \textit{Gaia} DR2 \citep{GaiaCollaboration_2018}. 
\texttt{age\textunderscore lowess\textunderscore correct} is a measurement of the stellar age from a neural network trained on \(\num{6676}\) stars with both spectroscopic measurement by APOGEE DR14 and asteroseismic age measurement by the \textit{Kepler} mission \citep{Borucki_2010}, corrected for biases from the neural network as described in \cite{Mackereth_2017}. 
Due to a lack of stars with low metallicity in the training data, reliable ages aren't available for stars with \(\FeH<-0.5\), so for these stars we must assume an age distribution as described below. 

The APOGEE DR16 ``statistical sample" (the sample of stars for which the selection function can be reconstructed) contains 165,768 stars.
However, this is partly made up of dwarfs, which have higher uncertainties in their atmospheric parameters and abundances. 
We thus select a subset of the statistical sample with a calibrated ASPCAP surface gravity of \(1\leq\log g<3\).
We additionally select only stars with fractional uncertainty in heliocentric distance \(D\) of less than 0.5.
To restrict our sample to the Milky Way's disk, we select stars with Galactocentric radii \(R\) between \qty{4}{\kilo\parsec} and \qty{12}{\kilo\parsec} and height above the disk \(z\) of \qty{-5}{\kilo\parsec} to \qty{5}{\kilo\parsec}.
This gives us a sample of \(\num{80958}\) red giants.

\subsection{Density Modelling of Red Giants across the Galaxy}\label{sec:densitymodelling}
To calculate the distribution of red giant stars in the Milky Way from the APOGEE data we use the method of \cite{Bovy_2016b}, as this accounts for both dust and the survey selection function simultaneously. 
In brief, assuming the stars observed in a survey are distributed independently in a space of some observables \(O\) (for example position, colour and magnitude, chemical abundances), then the positions of \(N\) stars in the space of observables \(O_1, \ldots, O_N\) is a realisation of an inhomogeneous Poisson point process. 
This process is a random distribution of points defined by a rate function \(\lambda(O)\), such that the number of points in a given volume \(V\) in the space of the observables is a Poisson random variable with mean and variance equal to the integral of the rate function over that volume, \(\int_V \lambda(O)\odif{O}\). 
It follows that the probability of finding a point (i.e. an observed star) with observables in the infinitesimal volume \(\fdif{O}\) is given by \(\lambda(O)\fdif{O}\), and the total number of points (i.e. stars observed) is a Poisson random variable with mean and variance \(\Lambda =\int \lambda(O)\odif{O}\). 
Since the rate function is equal to the rate of occurrence of observed stars, it can account for both the underlying true density of stars, as well as the effect of the survey selection function and dust which prevent all extant stars from being observed. 
If the rate function is modelled with \(\lambda(O\mid\theta)\), where \(\theta\) parameterises the model, then the likelihood of the model given \(N\) observed stars \(O_1, \ldots, O_N\) is given by
\begin{equation}\label{eq:genLikelihood}
    \ln \mathcal{L}(\theta) = \sum_i \ln \lambda(O_i\mid\theta) - \int \lambda(O\mid\theta)\odif{O} \, .
\end{equation}

APOGEE has a selection function based on bins in dereddened colour and apparent magnitude, so \cite{Bovy_2016b} define the effective selection function \(\mathfrak{S}\), a convenient quantity equal to the fraction of a population's stars at each heliocentric distance \(D\) that will be spectroscopically observed in each of APOGEE's fields.
This is calculated for each field by placing a tracer sample of stars in the field at that distance, and calculating the fraction that would be observed, given a dust map and the survey selection function in dereddened colour and apparent magnitude.
We evaluate this for each monoabundance population in each field at a range of heliocentric distances \(D\) and ages \(\tau\). 
This allows us to treat stars as only having three observables: the field they appear in, their distance from the Sun \(D\), and their age \(\tau\).
We used the effective selection function implementation in the \texttt{apogee}\footnote{\url{https://github.com/jobovy/apogee}} package, described in \cite{Bovy_2016b}. 
We obtained the tracer population by sampling PARSEC stellar model isochrones \citep{Bressan_2012, Marigo_2017} at a range of ages with a Kroupa initial mass function with a minimum mass of \(0.08M_\odot\) \citep{Kroupa_2001}, cut to \(1\leq\log g<3\) to match the APOGEE red giant sample to which it was being fit.
For a dust map we used a combination of \cite{Drimmel_2003}, \cite{Marshall_2006} and \cite{Green_2019}, combined with the package \texttt{mwdust}\footnote{\url{https://github.com/jobovy/mwdust}}, also described in \cite{Bovy_2016b}. 

Following \cite{Bovy_2016a}, we separate our red giant sample into monoabundance populations by binning the stars in \(\FeH\) and \(\alphaFe\), then fit a separate number density model \(n_\text{giants}\) to each monoabundance population.
The density model we chose to fit to each monoabundance population was a simple axisymmetric exponential in Galactocentric radius \(R\) and height above the plane of the disk \(z\), 
\begin{equation}
n_\text{giants}(R,z\mid \text{logA}, a_R, a_z) = \exp(\text{logA} -a_R(R-R_0) -a_z\lvert z\rvert)\, \unit{\per\kilo\parsec\cubed}\, ,
\end{equation}
parameterised by an amplitude logA and two scale parameters \(a_R\) and \(a_z\). 
\(R_0=\qty{8.1}{\kilo\parsec}\) is the radial distance of the Sun from the Galactic centre \citep{GRAVITYCollaboration_2018}. 
Note that both \(R\) and \(z\) are functions of our observables, the angular coordinates of the field pointing and distance from the Sun \(D\).
We simultaneously fit the age distribution of each monoabundance population, assumed to be a normal distribution with mean \(\tau_0\) and variance \(1/\omega\), 

\begin{equation}
g(\tau\mid\tau_0,\omega) = \sqrt{\frac{\omega}{2\pi}}\exp\left(-\frac{\omega}{2}(\tau-\tau_0)^2\right)\, 
\end{equation}
This form is justified by the simple narrow, monomodal age distributions for monoabundance populations found by \cite{Lian_2022}, and the fact that the effective selection function and IMF are sufficiently well-behaved that small changes in the age distribution will not significantly change our results.

To calculate the rate function \(\lambda(\text{field},D, \tau)\) for each monoabundance population, the number density of red giants needs to be multiplied a Jacobian factor \(\lvert J(\text{field}, D)\rvert = \Omega_\text{field}D^2\) to convert it from a density in Cartesian coordinates to a density in \(D\). 
Multiplying by the age distribution gives the combined underlying distribution in field, distance and age. 
The observed distribution is then found by multiplying this underlying distribution by the effective selection function \(\mathfrak{S}(\text{field},D, \tau)\). Thus the rate function for each monoabundance population is given by
\begin{equation}
\begin{split}
\lambda(\text{field},D,\tau \mid \text{logA}, a_R, a_z, \tau_0, \omega) &= n_\text{giants}(R,z\mid \text{logA}, a_R, a_z) \cdot g(\tau\mid\tau_0,\omega)\\
&\quad \cdot \lvert J(\text{field},D)\rvert\cdot \mathfrak{S}(\text{field},D,\tau) \, .
\end{split}
\end{equation}

\begin{figure}[h]
\newcommand{\fitfigwidth}{0.48}
\centering
\includegraphics[width=\fitfigwidth\textwidth]{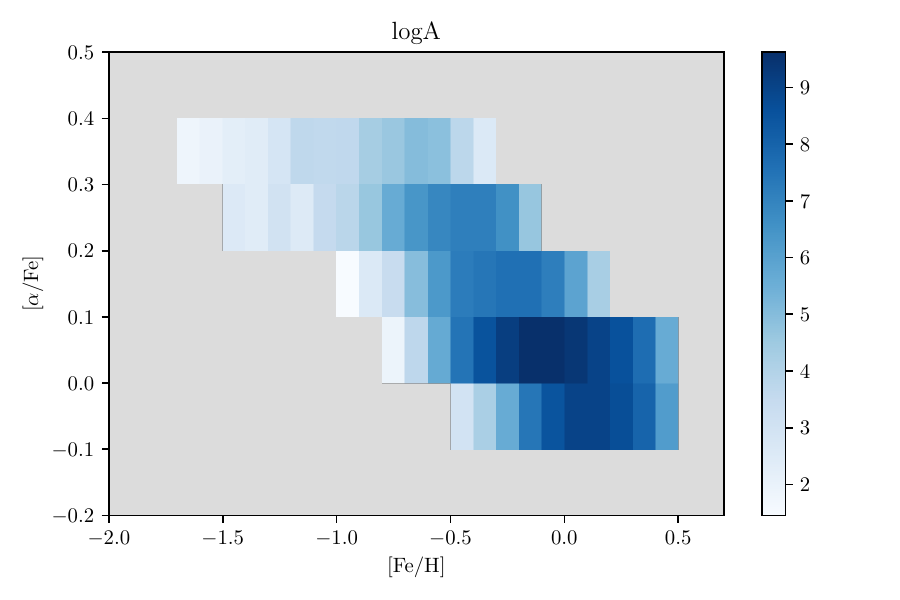}
\includegraphics[width=\fitfigwidth\textwidth]{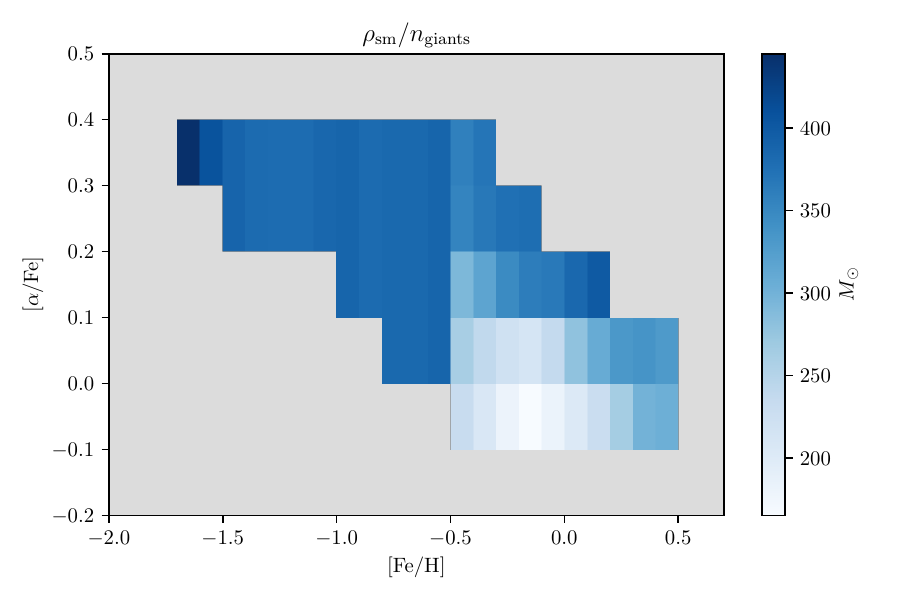}

\includegraphics[width=\fitfigwidth\textwidth]{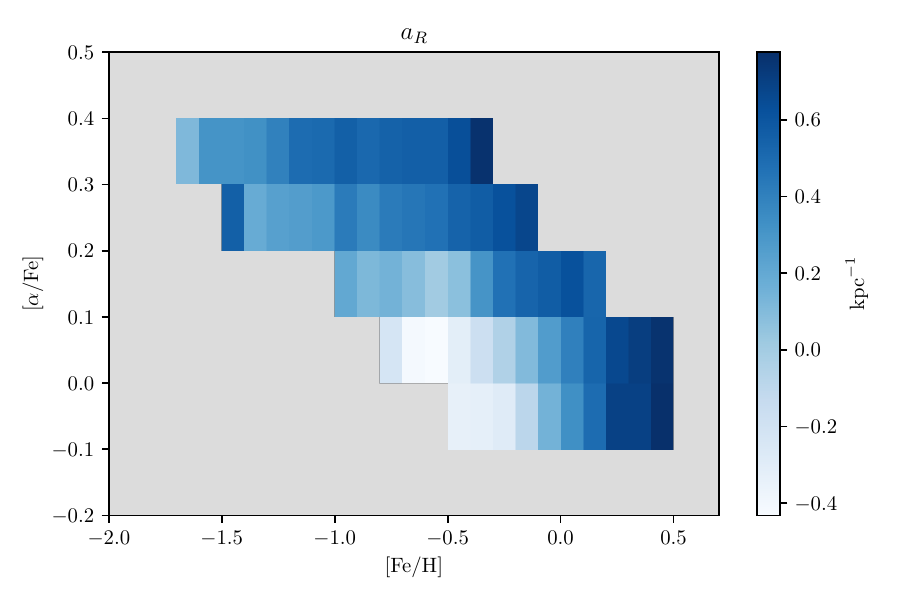}
\includegraphics[width=\fitfigwidth\textwidth]{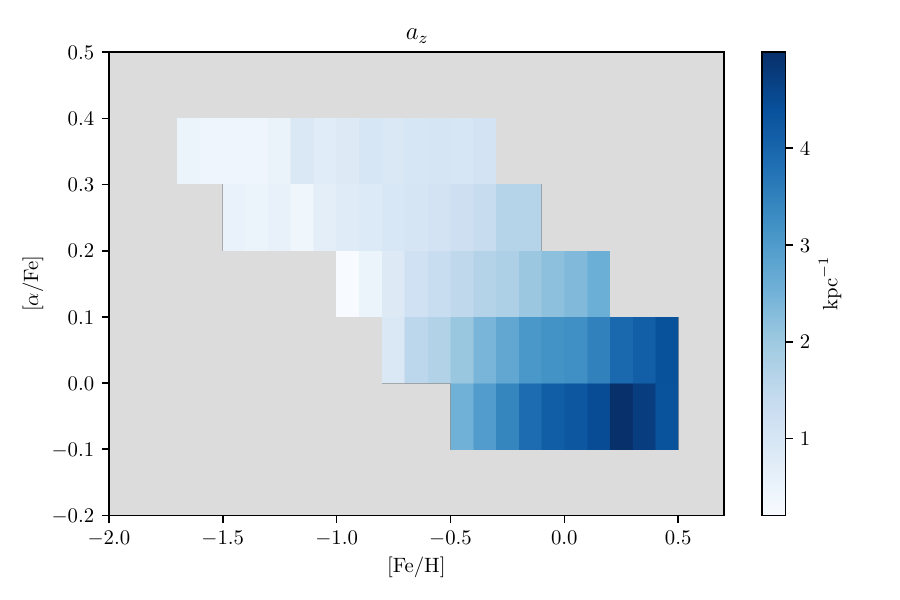}

\includegraphics[width=\fitfigwidth\textwidth]{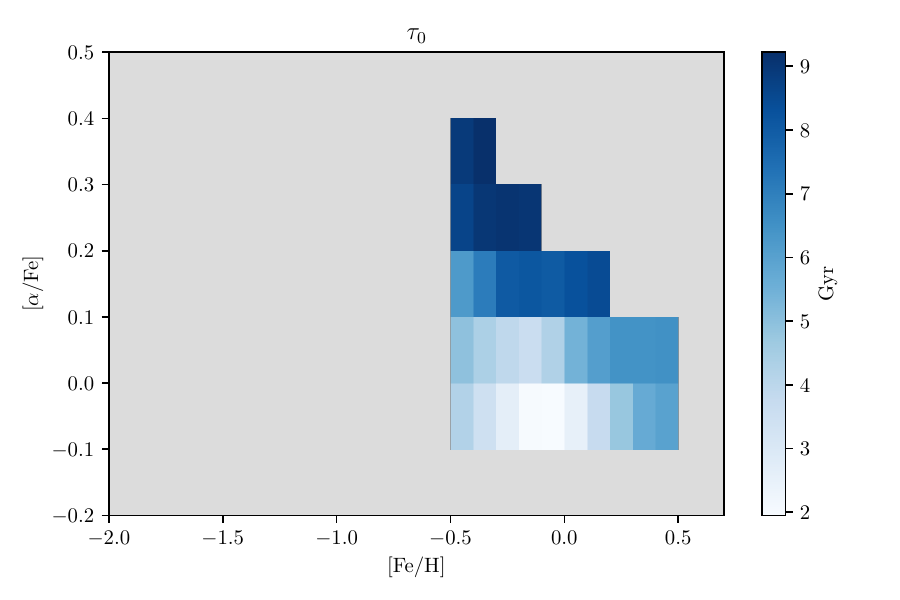}
\includegraphics[width=\fitfigwidth\textwidth]{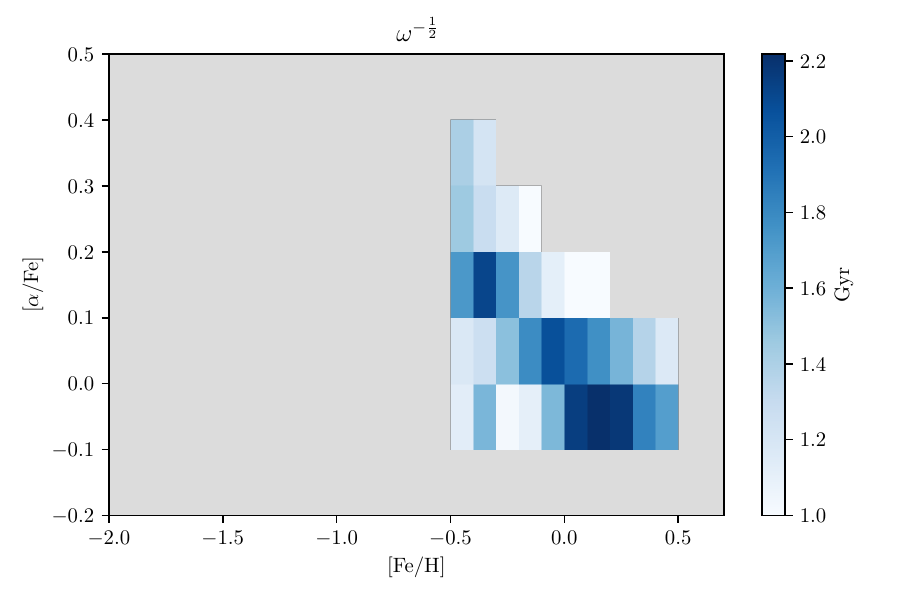}

\caption{The best-fit values of the density modelling parameters for each monoabundance population with 20 or more observed stars, and \(\rho_{\text{sm}}/n_{\text{giants}}\), the ratio between the \textit{sine morte} stellar mass density and the red giant number density, explained in section \ref{sec:sinemorte}.
}
	\label{fig:scales}
\end{figure}

This particular form for the density profile has the advantage that the Poisson point process likelihood takes the tractable form 
\begin{equation}
\begin{split}
\ln \mathcal L(\text{logA}, a_R, a_z, \tau_0, \omega) =\text{const} &+ N\left(\text{logA} - a_R\langle R -R_0\rangle -a_z\langle\lvert z\rvert\rangle +\tfrac{1}{2}\ln \omega -\frac{\omega}{2}\left(\langle\tau^2\rangle - 2\tau_0\langle\tau\rangle+\tau_0^2\right)\right)\\
&- \sum_{\text{field}}\int \int\lambda(\text{field},D\mid \text{logA}, a_R, a_z, \tau_0, \omega) \odif{D}\odif{\tau} \, ,
\end{split}
\end{equation}
where the sum over every star in the monoabundance population \(\sum_i \ln \lambda(O_i\mid\theta) \) in Eq.~\ref{eq:genLikelihood} is reduced to a linear combination of the parameters with aggregates of the data in the monoabundance population being fitted: \(N\) as the number of stars observed, \(\langle R\rangle\) and \(\langle\lvert z\rvert\rangle\) as the mean coordinate values, and \(\langle \tau\rangle\) and \(\langle \tau^2\rangle\) as the mean age and age squared. 
To build our model of the Milky Way disk between \(R=\qty{4}{\kilo\parsec}-\qty{12}{\kilo\parsec}\) and \(\lvert z\rvert=0-\qty{5}{\kilo\parsec}\), we found a best-fit model for each monoabundance population by maximising this likelihood with respect to logA, \(a_R\), \(a_z\), \(\tau_0\), and \(\omega\), visually checking each fit to ensure the global maximum had been found.
The results are plotted in Figure~\ref{fig:scales}.

Reliable ages are not available for stars \(\FeH<-0.5\) in \texttt{AstroNN}, so for these populations we fit only for logA, \(a_R\) and \(a_z\). 
We use an effective selection function dependent only on field and \(D\), calculated assuming a uniform age distribution. 
While a uniform age distribution is an inaccurate description of a monoabundance population, it is a non-informative assumption that ensures that the effective selection function is never zero where a star may actually be observed. 
On testing, we found that the effective selection function did not vary strongly with the age distribution assumed. 
Even then, as discussed in the section below, we do not expect stars with metallicity \(\FeH<-0.5\) to contribute a significant number of ISOs, making our conclusions independent of the age distribution assumed for these stars.
 
The two main distinct chemodynamical populations of the Milky Way are clearly shown in our model (Fig.~\ref{fig:scales}). 
The young, high \(\FeH\), low \(\alphaFe\) monoabundance populations have the low vertical scale lengths and high radial scale lengths which form the thin disk, whereas the old low \(\FeH\), high \(\alphaFe\) monoabundance populations have the opposite trend in scale lengths, forming the thick disk \citep{Mashonkina_2019}. 

This approach gives us simple but accurate models of the trends of the Milky Way disk's stellar population. These results agree with the results of \cite{Bovy_2016a} which also models monoabundance populations in APOGEE, fitting more complex models to only stars in the red clump, using their highly consistent absolute magnitudes as an accurate distance measurement.

\subsection{The \textit{Sine Morte} Stellar Population}
\label{sec:sinemorte}
Having obtained a model for the distribution of red giants in the Milky Way, we then infer the distribution of all stars throughout Galactic history. 
As described in \S~\ref{sec:stars2ISOs}, ISOs form mostly at the start of their parent star's life and then outlive their parent star. 
Under this constraint, the population of currently living stars is too limited to use to predict the ISO population. 
Instead, we must consider what the stellar population would be at the present time if stars did not die, but instead continued to orbit around the Milky Way indefinitely --- while being subjected to the same dynamical effects that affected both their longer-lived companion stars and the ISOs they had released on similar orbits. 
We introduce this as the \textit{sine morte}\footnote{Latin for ``without death'', [\textprimstress si\textlengthmark ne \textprimstress m\textopeno rte] IPA pronunciation, or `seen-ay mort-ay'.} stellar population. 

First, we calculate the \textit{sine morte} number density of stars in each monoabundance population, from the red giant number density. 
For this, we use the same PARSEC stellar model isochrones and Kroupa initial mass function with a minimum mass of \(0.08M_\odot\) as in \S~\ref{sec:data}. 
Using the isochrones with metallicities corresponding to each monoabundance population, weighted by the fitted age distribution \(g(\tau\mid\tau_0,\omega)\) for that monoabundance population, we calculate \(N_\text{giants}/N_\text{sm}\): the fraction of all stars ever created that are currently in the red giant phase, which is again defined by \(1\leq\log g<3\). 
Dividing the number density of giants by this fraction gives us the \textit{sine morte} stellar number density. This assumes that the age distribution of each monoabundance population does not vary significantly across the Milky Way disk, which is reasonable for the range of \(R\) and \(z\) we are considering \citep{Lian_2022}. 

\begin{figure}[h]
	\centering
    \includegraphics[width=\textwidth]{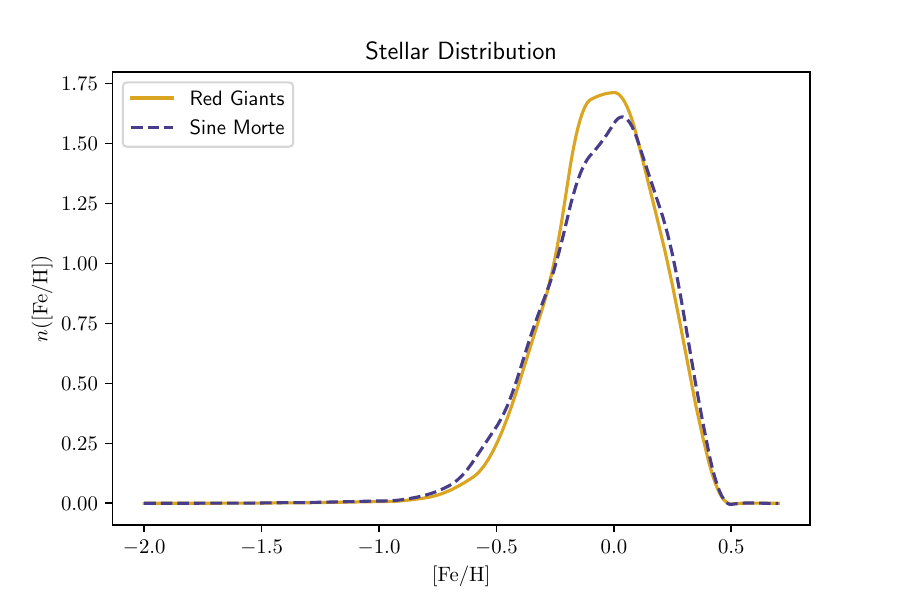}
	\caption{Comparison of the \(\FeH\) distribution of red giants and the \(\FeH\) distribution of the \textit{sine morte} stellar population around the Sun, both normalised. Though similar, there are subtle differences largely caused by the fact that the distribution of ages changes as a function of \(\FeH\), as shown in Fig.~\ref{fig:scales}.}
	\label{fig:NRGvsSMM}
\end{figure}

Next, we calculate the \textit{sine morte} stellar mass density by multiplying the \textit{sine morte} stellar number density by the average star's initial mass, \(\langle M_\text{int} \rangle\). 
By definition, the average mass of stars in the \textit{sine morte} population is the average initial mass, which is only dependent on the initial mass function. 
Thus we calculate \(\langle M_\text{int} \rangle\) from the same Kroupa IMF. 
All combined, the \textit{sine morte} stellar mass density is given by
\begin{equation}\label{eq:rhoSM}
\rho_\text{sm} (\mathbf{x}) = \frac{\langle M_\text{int}\rangle}{N_\text{giants}/N_\text{{sm}}} \cdot n_\text{giants}(\mathbf{x}) \, .
\end{equation}
The difference between the two populations at the position of the Sun is illustrated in Fig.~\ref{fig:NRGvsSMM}. 
The ratio between the \textit{sine morte} stellar mass density and the red giant number density is plotted in the top right panel in Fig.~\ref{fig:scales}.

For the monoabundance populations with \(\FeH<-0.5\), without reliable age measurements, we calculate an upper limit for the \textit{sine morte} mass density. 
We assume an old age distribution, with mean \(\tau_0=\qty{12}{\giga\year}\) and standard deviation \(\omega^{-\frac{1}{2}}=\qty{1}{\giga\year}\), which minimises \(N_\text{giants}/N_\text{sm}\). 
Even with this upper limit, below we find that stars with \(\FeH<-0.5\) contribute a very small number of ISOs, making our conclusions independent of this approximation.

\begin{figure}[h]
	\centering
    \includegraphics[width=\textwidth]{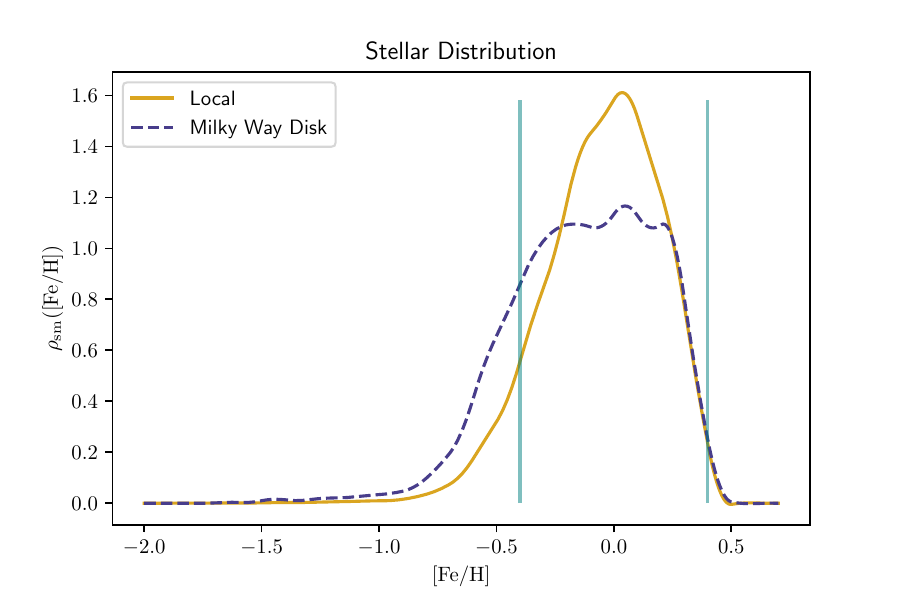}
	\caption{The normalised mass-weighted \textit{sine morte} stellar metallicity distribution, evaluated at the position of the Sun and integrated over the range of the Milky Way disk we are modelling. Vertical lines show the range in \(\FeH\) for which we model variations in the water mass fraction \(\fHHO\) of ISOs}
	\label{fig:FeH}
\end{figure}

The chemical model we use in \S~\ref{sec:PPD} to link the composition of ISOs to the composition of stars depends only on stellar metallicity \(\FeH\): so when we evaluate the model we sum the \textit{sine morte} distribution over the bins in \(\alphaFe\) to get a distribution in only \(\FeH\) bins. 
To ensure accurately fitted models, we include only monoabundance populations with 20 or more observed stars. 
We then smooth this binned distribution by taking the derivative of a spline fit to the cumulative distribution, knotted at the edges of the bins. 
Plotted in Fig.~\ref{fig:FeH} is the \textit{sine morte} metallicity distribution \(\rho_\text{sm}(\FeH)\) evaluated at the position of the Sun (\(R=\qty{8.1}{\kilo\parsec}\), \(z=\qty{0.021}{\kilo\parsec}\), \cite{GRAVITYCollaboration_2018, Bennett_2019}), and integrated over the broader \(R=\qty{4}{\kilo\parsec}-\qty{12}{\kilo\parsec}\) and \(\lvert z\rvert=0-\qty{5}{\kilo\parsec}\) range of the Galactic disk we are modelling.

\section{Predicting the Interstellar Object Distribution}\label{sec:predict}
In the previous section, we calculated the \textit{sine morte} stellar population of the Milky Way from the APOGEE survey. 
In this section we describe how to predict an example physical property of the Galactic ISO population --- the ISO water mass fraction distribution --- from this stellar population.

\subsection{Protoplanetary Disk Model}\label{sec:PPD}
We make the foundational assertion that all ISOs we consider form as planetesimals in a protoplanetary disk \citep{'OumuamuaISSITeam_2019}. 
A protoplanetary disk has to first order the same composition as the star it forms around, since they both form from the same molecular cloud core. 
Under this assumption, \cite{Bitsch_2020} predict the composition of planetesimals formed around stars of different metallicities. 
They do this for stars with metallicities in the range \(-0.4\leq \FeH\leq0.4\), using the average elemental composition of stars at each value of \(\FeH\) in the GALAH DR2 \citep{Buder_2018}, and for planetesimals that form both interior and exterior to the water ice line: the inner edge of the region of the protoplanetary disk where it is cool enough to form water ice. 
We assume that each star in the \textit{sine morte} population produces only ISOs with their composition set by the \cite{Bitsch_2020} formula for that metallicity, exterior to the water ice line. 
While in reality, stars will each produce a distribution of ISOs that formed at different positions in their protoplanetary disk and thus have a range of compositions, this simplification of only modelling planetesimals which form exterior to the water ice line is justified by the proportionally greater reservoir of snowline-exterior planetesimals, and the higher efficiencies of formation mechanisms dynamically stripping them into the interstellar population \citep{Fitzsimmons_2023}. 
In our Solar System, the vast majority of Oort cloud objects are ice-rich \citep{Meech_2016}; therefore both these and the majority of ISOs produced by the Solar System must have formed outside of the water ice line \citep{Filacchione_2022}. 
Additionally, planetesimals beyond the water ice line are more loosely bound to their parent stars, so will be more easily ejected (e.g. \cite{Moro-Martin_2018}). 

We focus on the mass fraction of water, \(\fHHO\), as this varies significantly and decreases monotonically with \(\FeH\) in the models of \cite{Bitsch_2020}. 
To obtain a smooth map from \(\FeH\) to \(\fHHO\), we fit a third-order polynomial to the water ice mass fraction data points in figure 10 of \cite{Bitsch_2020}. 
For metallicities outside the range of \(-0.4\leq\FeH\leq0.4\) we assume that the relationship between \(\fHHO\) and \(\FeH\) continues to be monotonic, with \(\fHHO\) remaining high beyond the low \(\FeH\) limit and remaining low beyond the high \(\FeH\) limit: allowing us to track the fraction of ISOs with high, low, and intermediate water mass fraction. 
This range corresponds to \(0.07\leq\fHHO\leq0.51\) in water mass fraction.

During the writing of this paper, \cite{Cabral_2023} was published, containing an updated version of the \cite{Bitsch_2020} chemical model which included data from GALAH DR3 \citep{Buder_2021} and APOGEE DR17 \citep{Abdurro'uf_2022}, and allowed for variation in \(\alphaFe\) as well as \(\FeH\). 
They found that the water mass fraction of planetesimals, consistent with the \cite{Bitsch_2020} assumptions we use, is most dependent on \(\FeH\). 
They also confirmed the trend between \(\FeH\) and \(\fHHO\) found in \cite{Bitsch_2020}.
However, they also found that for the APOGEE survey there was a smaller variation in \(\fHHO\) over the same range in \(\FeH\) compared to the GALAH data used in their work. 
\cite{Cabral_2023} notes that this means that the trends seen are robust. 
The findings of \cite{Cabral_2023} do mean that the predictions of the ISO distributions in our work may overestimate the width of the distribution in \(\fHHO\).

We assume that every star produces ISOs, and the number of ISOs produced by each star depends on its mass and metallicity. \cite{Lu_2020} argue that the mass of planet-forming material in a protoplanetary disk is proportional to both the mass of the host star mass \(M_\ast\) and its metal mass fraction \(Z\) --- well approximated by \(Z_\odot 10^{\FeH}\) for small values of Z (\(Z_\odot= 0.0153\), \cite{Caffau_2011}). In the absence of confirmed and comprehensive knowledge of ISO formation mechanisms, we use this as a reasonable proxy for the number of ISOs produced by each star. 
However, the number of ISOs produced by a star may not be simply proportional to the mass of planet-forming material, because ISO production also requires the ejection of planetesimals --- which is dependent on system dynamical architecture.
One major ISO ejection pathway is scattering by giant planets, the occurrence of which has its own metallicity dependence \citep{Osborn_2020}. 
Additionally, scattering by giant planets may form ISOs with extra fragmentation, reweighting the number distribution towards lower masses and increasing the number of ISOs produced from the same mass of planet-forming material \citep{Raymond_2018a}. 

We therefore assume the number of ISOs produced by each star is proportional to its mass, while incorporating into the model a power law dependence on metallicity mass fraction: thus the number of ISOs produced is proportional to \(M_* \cdot 10^{\beta\FeH}\). 
Here \(\beta=1\) corresponds to the simple assumption that the number of ISOs produced is proportional to the mass of planet-forming material. 
\(\beta=0\) corresponds to no metallicity dependence at all.
However, we expect \(\beta>0\): planetesimals require some fraction of dust and ice in order to exist, i.e. at the elemental level, C/N/O, Al/Si, \&c. must be present, so a minimum metallicity constraint must exist \citep{Johnson_2012}. 
We do not model the constant of proportionality here, as this depends on the size distribution of ISOs, which remains observationally unconstrained with only two ISOs \citep{'OumuamuaISSITeam_2019,Jewitt_2023}.
We predict only the normalised distribution of ISO water mass fractions.

\subsection{Predicting the ISO Population From the Stellar Population}\label{sec:stars2ISOs}

Since the majority of ISOs are expected to be ejected by dynamical processes within several hundred Myr of a star's formation \citep{Pfalzner_2019, Fitzsimmons_2023}, we assume that the birth of a star and release of its ISOs are contemporaneous, and omit accounting for any delay. 
A fraction of a planetary system's bound planetesimals will become unbound later in a system's life, whether from its Oort cloud's continuous interaction with the Galactic environment, or in post-main-sequence escape \citep{Veras_2011,Veras_2014,Levine_2023}; for simplicity we omit this smaller population here.
As Oort cloud comets, which also occupy an interstellar environment \citep{Kaib_2022}, do not exhibit substantive erosion after 4.5 Gyr \citep{Stern_1988}, we continue with the expectation that ISO erosional processes are broadly similar to those of Solar System comets \citep[e.g.][]{Guilbert-Lepoutre_2015}. 
This implies ISOs from stars prior to the Sun will outlive their parent stars.
We thus use the \textit{sine morte} stellar distribution established in \S~\ref{sec:sinemorte}. 
For assessing the water mass fraction of the resulting population, we assume processing in the interstellar medium has a negligible effect\footnote{While this is valid for water, it may not hold for hypervolatiles; e.g. \cite{Seligman_2022} suggest that processing in the interstellar medium will remove CO and C\(\mathrm{O}_2\) relative to \(\mathrm{H}_2\)O. However, this is countered by the CO-rich nature of 2I.}, and that ISOs broadly represent the water mass fraction of their source planetesimal populations.

To link the spatial distribution of stars throughout the Galaxy to the ISOs they produced, we need to model their combined Galactic dynamics.
ISOs are not expected to remain near to their parent stars on Gyr timescales, and stars do die. 
However, it is helpful to consider what the behaviour of the ISO population is in the case that neither of these are true. 
If ISOs did remain near their parent stars and both ISOs and stars existed for an infinite length of time, the number density distribution of ISOs with a given water mass fraction \(\fHHO\) and position in the Galaxy \(\mathbf{x}\) in the present day, denoted \(n_\text{ISO} (\mathbf{x}, \fHHO)\), would be equal to the number of ISOs produced by stars currently at the same position and of the corresponding metallicity \(\FeH\). 
Under the model of \S~\ref{sec:PPD}, the number of ISOs produced by a star of mass \(M_*\) and metallicity \(\FeH\) is proportional to \(M_* \cdot 10^{\beta\FeH}\), thus
\begin{equation}\label{eq:stayswith}
n_\text{ISO}(\mathbf{x}, \fHHO \mid \beta) \propto 10^{\beta\FeH} \cdot \odv{\FeH}{\fHHO} \cdot \rho(\mathbf{x}, \FeH) 
\end{equation}
where \(\rho(\mathbf{x}, \FeH)\) is the mass density distribution of stars at position \(\mathbf{x}\) with the metallicity \(\FeH\) corresponding to the ISO water mass fraction \(\fHHO\), and \(\odif{\FeH}/\odif{\fHHO}\) is the gradient of the relationship between \(\FeH\) and \(\fHHO\) described in \S~\ref{sec:PPD}.

The fact that stars do in fact die is then corrected for by replacing \(\rho(\mathbf{x}, \FeH)\) in this equation with \(\rho_{\text{sm}}(\mathbf{x}, \FeH)\), the \textit{sine morte} stellar mass density introduced in \S~\ref{sec:sinemorte}. 
Correcting for the fact that ISOs disperse from near their parent stars is more complex, so we proceed with some simplifying assumptions.
Once ejected, unless its resultant velocity exceeds the Galactic escape velocity, an ISO will orbit the Galactic centre, as stars do. 
We assume ISOs are ejected relatively slowly from their parent planetary systems compared to the stellar velocity dispersion. 
This is justified assuming ejection velocities \(<\qty{10}{\kilo\metre\per\second}\), the maximum ejection velocity from a planetary system under an expected suite of scattering mechanisms \citep{Pfalzner_2019, Fitzsimmons_2023}, and the velocity dispersions \(\gtrsim\qty{20}{\kilo\metre\per\second}\) measured in the Solar Neighbourhood \citep{Anguiano_2020}. 
Stars form on near-circular orbits around the Galactic centre \citep{Frankel_2020}, so a cloud of recently ejected ISOs will all have similar orbits to their parent star: nearly circular with similar ranges of oscillation in Galactocentric \(R\) and \(z\). 
However, the slight differences in their orbits will give the ISOs different orbital periods around the Galactic centre, meaning they will disperse along their similar near-circular orbital paths. 
Therefore, though ISOs do not stay near their parent star, we assume here that they only disperse in the azimuthal direction, and remain in the same \(R\) and \(z\) range as their parent star. 
Under this assumption equation \ref{eq:stayswith} still holds if the stellar density model is axisymmetric, depending only on \(R\) and \(z\), as does ours:
\begin{equation}\label{eq:theEq}
n_\text{ISO}(R,z, \fHHO \mid \beta) \propto 10^{\beta\FeH} \cdot \odv{\FeH}{\fHHO} \cdot \rho_{\text{sm}}(R,z, \FeH)
\end{equation}

 Orbits around the Galactic centre can evolve with time, due to the influence of perturbing potentials such as spiral arms, the bar and molecular clouds. 
 These effects can cause both dynamical `heating', an increase in the size of radial and vertical excursions from a circular orbit, and `migration', changes to the radius of an orbit while it remains nearly circular \citep{Sellwood_2002}.
 We introduced Eq.~\ref{eq:theEq} with the assumption that an ISO will stay in the same range of \(R\) and \(z\) as its parent star.
 Due to their azimuthal separation, the star and ISOs will experience slightly different perturbing potentials, causing their orbits to evolve adjacently but independently. 
 However, if the Galactic stellar and ISO distributions are sufficiently axisymmetric, perturbations will consistently change together both the orbits of stars and the orbits of a corresponding number of ISOs. 
Thus, Eq.~\ref{eq:theEq} holds for our model.

In this work we predict the distribution of ISOs in \(\fHHO\) both at particular values of \(R\) and \(z\) and integrated over the whole Milky Way. 
Since we do not model the total number of ISOs, we remove the need for the constant of proportionality in Eq.~\ref{eq:theEq} by normalising each \(n_\text{ISO}\) distribution we calculate with
\begin{equation}
p(\fHHO\mid \beta) = \frac{n_{\mathrm{ISO}}(\fHHO\mid \beta)}{\int n_{\mathrm{ISO}}(\fHHO\mid \beta) \odif{\fHHO}} \, .
\end{equation}

This gives us the distribution of ISOs within the bounds of the protoplanetary disk chemical model: \(0.07\leq\fHHO\leq0.51\). 
Outside of this range, we can calculate the fraction of ISOs with \(\fHHO\leq0.07\) and \(\fHHO\geq0.51\), by assuming that the relation between \(\FeH\) and \(\fHHO\) remains monotonic.
Thus, all stars with \(\FeH\leq-0.4\) contribute ISOs with \(\fHHO\geq0.51\), and all stars with \(\FeH\geq0.4\) contribute ISOs with \(\fHHO\leq0.07\).

\section{Results}\label{sec:results}
In this section we demonstrate the prediction framework of section \ref{sec:predict} by making two different predictions.
We demonstrate two different example values for a stellar metallicity dependence for ISO production, \(\beta\), with \(\beta=1\) for our principal prediction and \(\beta=0\) as an alternate prediction.

\subsection{Principal Prediction: \texorpdfstring{\(\beta=1\)}{beta=1}}\label{sec:mainRes}
First, we predict the distribution of ISOs assuming that the number produced by each star is proportional to the star's metal mass fraction \(Z\), by setting \(\beta=1\) in equation \ref{eq:theEq}. 
As described in section \ref{sec:PPD}, in the absence of concrete knowledge of ISO formation mechanisms this is a reasonable value to assume, and thus we consider
this our principal prediction.

\begin{table}[h]
\centering
\begin{tabular}{ccc} 
 \tableline
ISO $\fHHO$ range & Fraction of ISOs around Sun & Fraction of ISOs in Milky Way Disk\\ \hline
$\fHHO<0.07$ & 0.017 & 0.024 \\
 $0.07\leq\fHHO\leq0.51$ & 0.955 & 0.915 \\ 
 $0.51<\fHHO$ & 0.027 & 0.061 \\
 \tableline
\end{tabular}
\caption{Primary prediction for the fraction of ISOs in each \(\fHHO\) range evaluated at the position of the Sun and integrated over whole Milky Way, with \(\beta=1\).}
\label{tab:beta1}
\end{table}

Table \ref{tab:beta1} lists the the fraction of ISOs within and outside either end of the \cite{Bitsch_2020} protoplanetary disk chemical model \(\fHHO\) range, $0.07\leq\fHHO\leq0.51$. 
We assess both the distribution of ISOs at the position of the Sun, at \(R=\qty{8.1}{\kilo\parsec}\), \(z=\qty{0.021}{\kilo\parsec}\) \citep{GRAVITYCollaboration_2018, Bennett_2019}, and the distribution of ISOs integrated over the region of the Milky Way disk we are modelling. 
In both cases, the vast majority of ISOs lie within the range of the model. 
The significant mass of stars beyond the lower \(\FeH\) limit of the \(\rho_\text{sm}(\FeH)\) distribution in Fig.~\ref{fig:FeH}, both at the position of the Sun and over the whole Milky Way disk, do not contribute significantly to the fractions of ISOs in the high \(\fHHO\) range.
This is because of the exponential dependence of the number of ISOs produced by each star on \(\FeH\) in Eq.~\ref{eq:theEq}.

\begin{figure}[h]
	\centering
    \includegraphics[width=\textwidth]{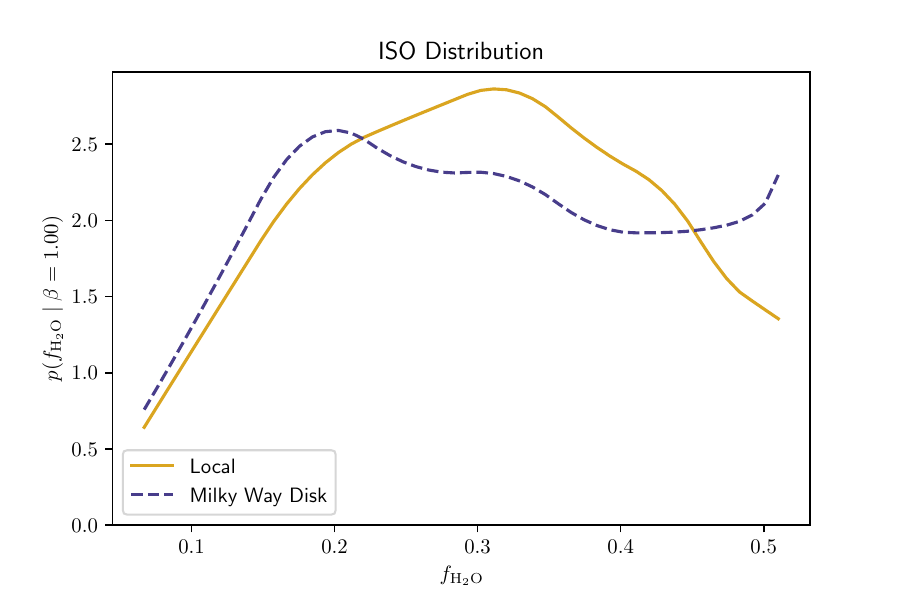}
	\caption{Primary prediction for the distribution of ISO water mass fractions, evaluated at the position of the Sun and integrated over the Milky Way disk, for \(\beta=1\).}
	\label{fig:beta1}
\end{figure}

Within the range of the chemical model, we can plot the distribution function of the ISO water mass fractions. 
Figure~\ref{fig:beta1} shows both the population of ISOs around the Sun and over the whole Milky Way. 
The different shapes of the two metallicity distribution functions in Figure~\ref{fig:FeH} is also apparent here, as the wider Milky Way metallicity distribution function results in a wider ISO water mass fraction distribution.

\begin{figure}[h]
\newcommand{\Rfigwidth}{0.49}
        \centering
        \includegraphics[width=\Rfigwidth\textwidth]{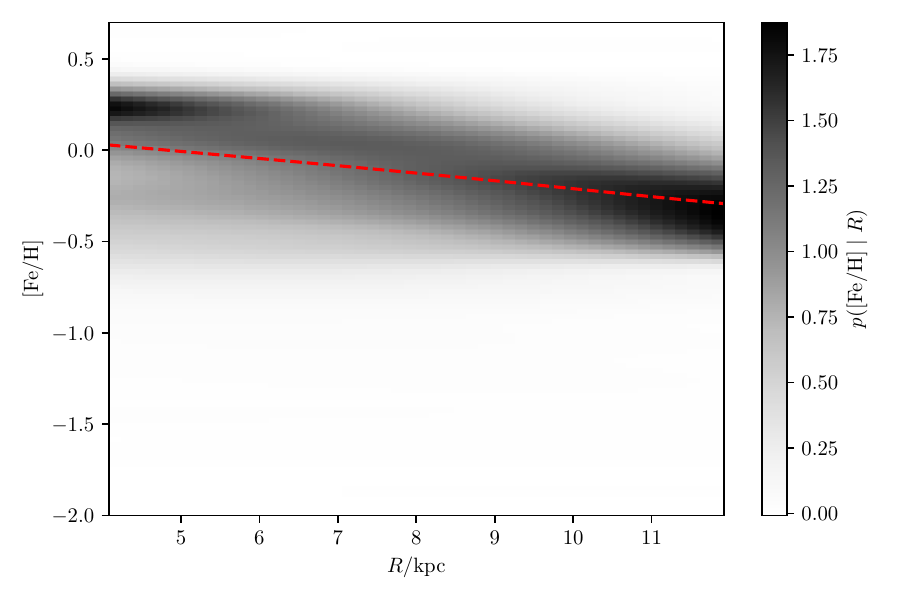}
        \includegraphics[width=\Rfigwidth\textwidth]{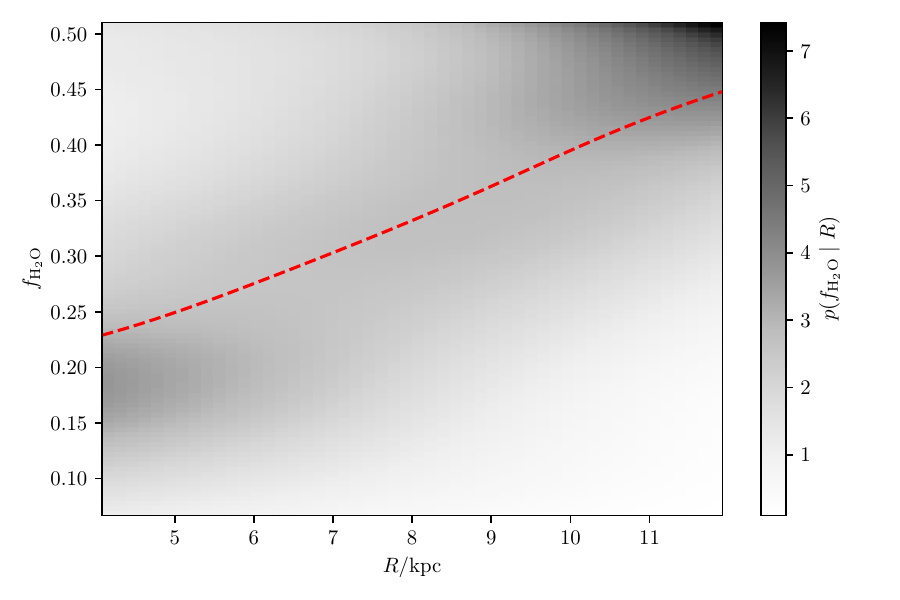}
	\caption{\textit{sine morte} stellar mass distributions (left) and ISO water mass fraction distributions (right), integrated over \(z\), at each distance to the Galactic Centre \(R\). Both distributions are normalised such that at each value of \(R\) the integral over \(\FeH\) or \(\fHHO\) is unity. Dashed lines show the median value of each distribution at each \(R\).
 }
	\label{fig:DistsR}
\end{figure}

The distributions of ISOs around the Sun and averaged over the Milky Way disk are remarkably similar.
We explore this in Figure~\ref{fig:DistsR}, through the \textit{sine morte} stellar mass \(\FeH\) distribution and the ISO \(\fHHO\) distribution within the range of the chemical model at a range of values of \(R\), integrated over \(z\). 
Also plotted are the median values of \(\FeH\) and \(\fHHO\) at each value of \(R\). 
Clear in the left-hand panel of Fig.~\ref{fig:DistsR} is the well-studied Galactic metallicity gradient \citep{Cheng_2012} --- but additionally in the right-hand panel is a corresponding gradient in ISO water mass fraction. 
Since the composition of ISOs depends on the chemical makeup of the stars that they form around, we expect trends in the chemical abundances of stars to be accompanied by equivalent trends in the compositions of ISOs. 
Figure~\ref{fig:DistsR} also shows why the Solar neighbourhood distributions are similar to the whole-Galaxy integrated distributions in Figures \ref{fig:FeH} and \ref{fig:beta1}: The Solar neighbourhood happens to be at an intermediate value of \(R\) (\qty{8.1}{\kilo\parsec}), where both the stellar \(\FeH\) distribution and therefore the ISO \(\fHHO\) distributions are approximately midway between the high and low extremes.

In these plots, we have normalised the distributions in \(\FeH\) and \(\fHHO\) such that at each value of \(R\) the integral over \(\FeH\) or \(\fHHO\) is unity. 
However, it is worth noting that our model implies the densities of both the stellar and ISO populations will decrease with distance from the Galactic centre. 
The exponential stellar density profile is well established \citep{Juric_2008}.
Here we predict that the Galactic ISO population density profile will decrease faster than that of the stars: due to the metallicity dependence of the number of ISOs produced by each star, the higher-metallicity stars in the inner disk will produce more ISOs per unit of stellar mass than the lower-metallicity stars of the outer disk.

\subsection{Alternate Prediction: \texorpdfstring{\(\beta=0\)}{beta=0}}\label{sec:altRes}

\begin{figure}[h]
    \centering
    \includegraphics[width=\textwidth]{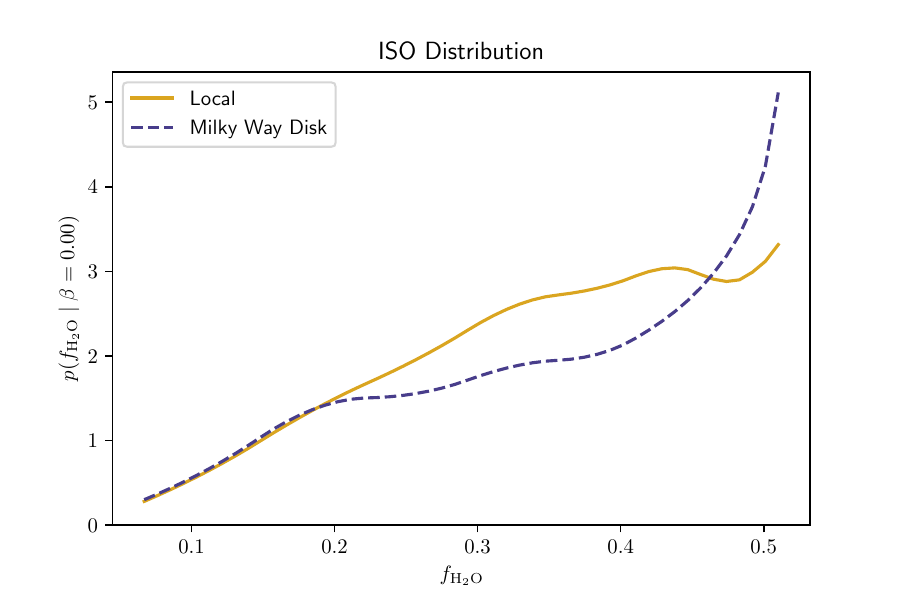}
	\caption{Alternate prediction for the distribution of ISO water mass fractions, evaluated at the position of the Sun and integrated over the Milky Way disk with \(\beta=0\): assuming the number of ISOs produced by each star is independent of the star's metallicity.}
	\label{fig:beta0}
\end{figure}

Setting \(\beta=1\) is just a choice in a basic and observationally unconstrained model. 
Therefore, we explore how changing its value affects the predicted ISO distribution. 
\cite{Lintott_2022} predicted the ISO populations of galaxies from the EAGLE hydrodynamical simulation using the same protoplanetary disk chemical model to map stellar metallicities \(\FeH\) to ISO water mass fractions \(\fHHO\).
However, \cite{Lintott_2022} assumed that the number of ISOs produced by each star was independent of the star's metallicity. This is equivalent to setting \(\beta=0\) in Eq.~\ref{eq:theEq} in this work, and we use this to make our alternate prediction. 
The resulting water mass fraction distributions for ISOs at the position of the Sun and integrated over the whole Milky Way are plotted in Figure~\ref{fig:beta0} and tabulated in Table~\ref{tab:beta0}.

\begin{table}[h]
\centering
\begin{tabular}{ccc} 
 \tableline
ISO $\fHHO$ range & Fraction of ISOs around Sun & Fraction of ISOs in Milky Way Disk\\ \hline
$\fHHO<0.07$ & 0.007 & 0.009 \\
 $0.07\leq\fHHO\leq0.51$ & 0.903 & 0.798 \\ 
 $0.51<\fHHO$ & 0.090 & 0.194 \\
 \tableline
\end{tabular}
\caption{Alternate prediction for fraction of ISOs in each \(\fHHO\) range, evaluated at the position of the Sun and integrated over the whole Milky Way with \(\beta=0\): assuming the number of ISOs produced by each star is independent of the star's metallicity.}
\label{tab:beta0}
\end{table}

Assuming \(\beta=0\) causes the ISO distribution to be more weighted towards a higher water mass fraction. 
These high water mass fraction ISOs come from low metallicity stars, which in this scenario no longer have their contributions to the ISO population suppressed by the exponential dependence on \(\FeH\). 
This means that the low metallicity tails of the \(\rho_\text{sm}(\FeH)\) distributions in Figure~\ref{fig:FeH} now contribute a significant number of ISOs with \(0.51<\fHHO\): almost 20\% when averaged over the Milky Way disk. 
A comparison of these results to those of \cite{Lintott_2022} is discussed in section \ref{sec:lintott2022}.

\section{Inference Framework}
\label{sec:inferenceFramework}\subsection{Bayesian Inference}\label{sec:Bayes}

In this work so far we have detailed a method of predicting the distribution of ISOs around the Sun, and demonstrated that this method can be used with varying models to get different predictions. 
These predictions can be compared to an observed sample of ISOs in order to make inferences about the models which went into producing them.
In this section, we detail how to do this in a Bayesian manner.

The predictions we make in this work are normalised ISO water mass fraction distributions, \(p(\fHHO\mid\beta)\), parameterised by the ISO production power law slope \(\beta\). This however could be generalised to a distribution in any set of observable properties \(\mathbf{O}\), and parameterised by any set of parameters \(\theta\). 
These distributions are also probability density functions for the value of \(\mathbf{O}\) of a randomly selected ISO, so \(p(\mathbf{O}\mid\theta) \fdif{\mathbf{O}}\) is the probability of an ISO having properties in the infinitesimal volume \(\fdif{\mathbf{O}}\). 
Extending this to multiple ISOs, a sample of ISOs with properties \({\mathbf{O}}_{1}, \ldots,{\mathbf{O}}_{N}\) will have likelihood
\[p({\mathbf{O}}_{1}, \ldots,{\mathbf{O}}_{N}\mid\theta) = \prod_i p({\mathbf{O}}_i\mid\theta)\]
assuming ISOs arrive independently with independent compositions. 
This is a valid assumption if the ISOs we observe are samples from a large Galactic population with contributions from stars across the Galaxy and cosmic time.
This is because the probability that two randomly selected ISOs have the same parent star is approximately equal to one over the total number of stars which could contribute ISOs to the population around the Sun.
The ISO sample likelihood can then be combined with priors to form a posterior distribution on the model parameters:
\begin{equation}\label{eq:posterior}
p(\theta\mid{\mathbf{O}}_{1}, \ldots,{\mathbf{O}}_{N}) \propto p(\theta)\cdot \prod_i p({\mathbf{O}}_i\mid\theta)
\end{equation}

This posterior can be used to calculate estimates and confidence intervals for the values of the parameters \(\theta\). An example of this calculation is given in section \ref{sec:betaInference}.

\subsection{Additional Properties of ISOs}\label{sec:Additions}

As with all minor planets, there are many observable quantities for ISOs that could be used in this framework in the set of observable properties \(\mathbf{O}\). 
If the distribution of ISOs in these properties can be predicted, the inference method of \S~\ref{sec:Bayes} can be used to compare the predicted distribution to the observed distribution in these properties, to make inferences about the models used. 
If multiple properties are included in \(\mathbf{O}\), then the joint distribution of ISOs in these properties can be predicted and used in inference. 
Including different ISO properties in the framework will allow inferences to be made about different processes affecting the ISO population.
We briefly consider several such properties here.

The Milky Way stellar population is broadly divided between different chemo-dynamical populations: the thin disk (high \(\FeH\), low velocity dispersion), thick disk (low \(\FeH\), high velocity dispersion), and the halo (very low \(\FeH\), radial orbits) \citep{Recio-Blanco_2014, Horta_2023}. 
Since the composition of an ISO depends on the metallicity of the star it formed around, each chemo-dynamical stellar population will contribute ISOs to the Galactic population with a distinct joint distribution in both composition and velocity. 
Therefore, including the velocities of ISOs in this framework could be used to tie ISOs to the chemo-dynamical stellar populations their parent star belongs to \citep{Eubanks_2021}. 
A prediction and discussion of the chemodynamical distribution of ISOs using stellar velocities from the \textit{Gaia} satellite is currently in preparation.

As an alternative measurement of composition to water mass fraction, the carbon-to-oxygen ratio of a cometary ISO can be estimated from its coma. 
This contains information about an ISO's formation location in the protoplanetary disk relative to the H\(_2\)O, CO\(_2\) and CO ice lines; modelling suggests it would be a useful measure for future ISOs \citep{Seligman_2022}. 

Further additions to our framework could include predictions of the size distribution and aspect ratio distribution of ISOs, which would be especially interesting due to 1I/\okina Oumuamua's extreme shape. 
The prediction of these distributions would depend on the stellar population via models of planetesimal formation which could link the size and shape distribution of planetesimals to the properties of their natal star. 
Finally, including the binarity rate of ISOs would test models of the applicability of formation mechanism of planetesimals to other protoplanetary disks, as the compact binarity of trans-Neptunian objects does for the Solar System protoplanetary disk \citep[e.g.][]{Nimmo_2018}. 
This could also constrain the ejection mechanisms of ISOs: loosely-bound wide planetesimal binaries would not survive scattering by a giant planet, but would survive a more gentle gravitational interaction with a stellar flyby.

\subsection{Generalisation to Other Galactic Populations}\label{sec:Generalisation}
Our method could easily be generalised to any Galaxy-wide population with a dependence on the properties of stars, simply by replacing the ``ISO recipe'' of section \ref{sec:predict}. 

For example, the distribution of planets through the Galaxy could be predicted by substituting a model of the occurrence rate of planets as a function of their host star's metallicity. 
This is based on the planet-metallicity correlation \citep{Fischer_2005, Osborn_2020}: as noted earlier that dust is necessary for planets, stars with higher metallicity are more likely to host planets. 
Here, the Milky Way metallicity gradient would mean that planets are more common towards the Galactic centre. 
This is a testable prediction with microlensing surveys such as OGLE \citep{Udalski_2015}, which continue to find exoplanets between the Solar System and the Galactic centre \citep{Tsapras_2018} with distances estimated from followup observations \citep{Vandorou_2023}.
If ISOs seed planet formation as hypothesised by \cite{Pfalzner_2019}, the ISO gradient we predict in \S~\ref{sec:mainRes} could also produce a signature in the planetary population.

Other variables affecting planet formation however could flatten or even reverse this gradient. Based on 28 planetary microlensing events, \cite{Koshimoto_2021} find tentative evidence of a decrease in the planet occurrence rate towards the Galactic centre, possibly due to the increased rate of close stellar encounters in the Galactic bulge.

\section{Discussion}
\label{sec:discussion}

\subsection{Comparison to Previous Work}\label{sec:lintott2022} 
As described in section \ref{sec:PPD}, \cite{Cabral_2023} updates the protoplanetary disk chemical model of \cite{Bitsch_2020} that we use, and note that the trend of planetesimal water mass fraction decreasing with stellar metallicity is robust. 
They do however find that for the APOGEE survey, there was a smaller variation in \(\fHHO\) over the same range in \(\FeH\) compared to the GALAH data used in \cite{Bitsch_2020}. 
If the true variation in \(\fHHO\) is smaller than in the model of this work, our predictions may overestimate the width of the ISO water mass fraction distribution.

\cite{Lintott_2022} made a prediction of the ISO population of a simulated Milky Way-like galaxy from the EAGLE hydrodynamical cosmological simulation \citep{Schaye_2015}, using a model equivalent to that of this work with \(\beta=0\). 
Whereas we predict a single-peaked \(\fHHO\) distribution, they predicted an ISO distribution with a significant number of ISOs with water mass fraction both below and above the range of the protoplanetary disk model, which they interpret as a bimodal distribution in ISO composition. 

There are expected reasons for the difference between the prediction here, based on the observed Milky Way stellar population, and the prediction by \cite{Lintott_2022} based on the simulated EAGLE galaxy. 
The EAGLE galaxy has a much wider \(\FeH\) distribution than the Milky Way, with many more stars outside of the \(\FeH\) range of the protoplanetary disk model. 
As a smoothed particle hydrodynamics simulation, EAGLE is susceptible to producing galaxies with \(\FeH\) distributions wider than those observed in nature, due to underestimating metal mixing between particles \citep{Wiersma_2009}. 
Therefore the results of this work, based on the observed stellar population of the Milky Way, should give a much more accurate prediction for the Milky Way's population of ISOs.

\subsection{Distinguishing Local and Galactic Populations of ISOs}

The results of \S~\ref{sec:mainRes} show that the well-studied metallicity gradient of the Milky Way has a corresponding ISO composition gradient, with ISOs generally having a higher water mass fraction at larger Galactocentric radii. 
Although we can only observe the compositions of ISOs which pass through the inner Solar System, it is still instructive to model how the distribution of ISOs varies across a wider portion of the Galactic disk.
This is because we have made assumptions about the Galactic dynamics of ISOs, under which the distribution of ISOs at a point in the Galaxy corresponds to the distribution of stars at that same point. If these assumptions break down, then the particular way in which they break down will affect the population of ISOs detectable in the Solar System in a related, calculable way.

For example, radial migration, caused by the non-axisymmetric potential of spiral arms, flattens the the Milky Way metallicity gradient by blurring the metallicity distribution in the radial direction \citep{Vickers_2021}. This widens the stellar metallicity distribution around the Sun as stars migrate in from the metal-poor outer disk and out from metal-rich inner disk. However, ISOs may undergo less radial migration than stars, due to the random motion given to them by their ejection from their home planetary system \citep{Daniel_2015}. 
The stars currently in the Solar neighbourhood may thus have a wider range of Galactocentric radii of origin than the ISOs. 
This would make the distribution of observable ISO compositions narrower than would be predicted from the distribution of stars. 

Additionally, the low velocity of 1I/\okina Oumuamua relative to the local standard of rest implies that it --- and therefore a large fraction of observable ISOs --- could come from local star-forming regions \citep[e.g.][]{Hallatt_2020}. 
This is also testable: their compositions will match those predicted from the metallicities of the nearest star forming regions. 
In addition, if the velocity distribution of ISOs is included in future work as described in \ref{sec:Additions}, it may be possible to trace individual ISOs back to the star forming regions they came from by matching them up with both composition and velocity. 
This would be hugely advantageous for studying planetesimal formation, as it would allow us to pair the properties of detected ISOs directly with observations of their parent planetary systems.

\subsection{An Estimation of the ISO Production Metallicity Dependence}\label{sec:betaInference}
The two different predictions made in \S~\ref{sec:results} demonstrate that the Galactic population of ISOs is sensitive to small changes to the processes that affect their formation and evolution. 
This means that if models of these processes can be combined to make accurate predictions of the ISO population, then the framework described in \S~\ref{sec:inferenceFramework} can be used to make inferences about those processes. 

In particular, the predictions of sections \ref{sec:mainRes} and \ref{sec:altRes} show that the ISO distribution around the Sun is sensitive to the metallicity dependence of the number of ISOs produced by each star, \(\beta\). 
As outlined in section \ref{sec:Bayes}, these two predictions can then be compared to a sample of ISOs.
In our example physical property, this is done by calculating the likelihood of each of these predictions producing the observed distribution of water mass fractions, such as for the ISOs expected to be found by the Legacy Survey of Space and Time (LSST) of the Vera C. Rubin Observatory.  
For a more general result, this likelihood can be combined with a prior on \(\beta\) and Bayes' theorem to calculate a posterior distribution for \(\beta\). 

Though the work of this paper has been carried out in expectation of a larger sample of ISOs being known in the future, we do already have a preliminary estimate of the ISO water mass fraction distribution. 
Due to the unknown composition of 1I/\okina Oumuamua, we can only estimate the water mass fraction of 2I/Borisov. 
We adopt a value of \(\fHHO=0.3\), after that of \cite{Seligman_2022} inferred from the production rates of 2I's coma. 
However, this value is more useful for demonstrative purposes than as a finely constrained estimate; estimating a comet's bulk composition from the composition of its coma is challenging. 
For instance, \cite{Seligman_2022} note that the compositions of interstellar comets calculated from production rates are affected by preferential desorbtion of CO and \(\textrm{CO}_2\) relative to \(\mathrm{H}_2\)O. 

\begin{figure}[h]
        \centering
        \includegraphics[width=\textwidth]{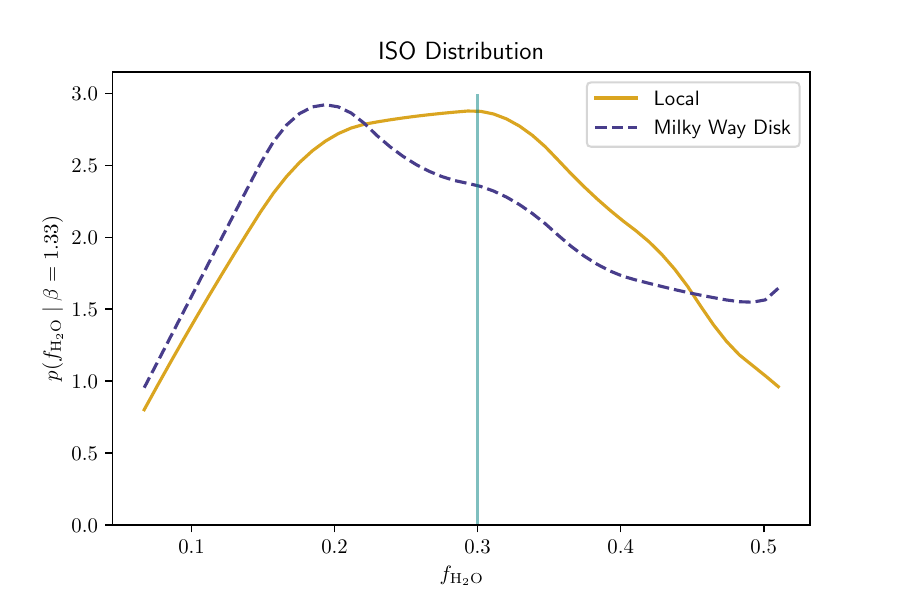}
	\caption{Distribution of ISO water mass fractions, evaluated at the position of the Sun and integrated over the Milky Way disk, with maximum \textit{a posteriori} estimate of \(\beta=1.33\). At \(\fHHO=0.3\) a vertical line marks the measured water mass fraction of 2I/Borisov.
 }
	\label{fig:optLocal}
\end{figure}

With this observed distribution of \(\fHHO\) we can then use the Bayesian framework of \S~\ref{sec:Bayes} to calculate a posterior distribution for \(\beta\).
We use our model for the distribution of ISOs around the Sun, and taking a uniform prior on \(\beta\) means that the posterior is simply proportional to the likelihood, equal to the value of the \(\fHHO\) distribution.
For \(\fHHO=0.3\) this posterior is maximised by a value of \(\beta=1.33\).
The distribution of ISOs at this value is listed and plotted in Table~\ref{tab:betaOpt} and Figure~\ref{fig:optLocal}. 
The symmetric 90\% confidence limit for this estimation of \(\beta\) is \(\beta\in(-1.3, 7.2)\). 
This is of course a wide interval containing physically implausible values.
The physically implausible values could be removed from the confidence interval by using a physically motivated prior on \(\beta\) --- but with one known value of \(\fHHO\) this would make the posterior dominated by the prior. 
Using the sample of ISOs that the LSST is expected to find will much better constrain the posterior for \(\beta\) and other parameters used in models with this framework. 

\begin{table}[h]
\centering
\begin{tabular}{ccc} 
 \tableline
ISO $\fHHO$ range & Fraction of ISOs around Sun & Fraction of ISOs in Milky Way Disk\\ \hline
$\fHHO<0.07$ & 0.022 & 0.031 \\
 $0.07\leq\fHHO\leq0.51$ & 0.960 & 0.930 \\ 
 $0.51<\fHHO$ & 0.018 & 0.040 \\
 \tableline
\end{tabular}
\caption{Predicted fraction of ISOs in each \(\fHHO\) range, evaluated at the position of the Sun and integrated over the Milky Way dist, with maximum \textit{a posteriori} estimate of \(\beta=1.33\).}
\label{tab:betaOpt}
\end{table}

\subsection{Selection Effects}
It should be noted that the predictions in this work ignore some specific effects which will influence the population of ISOs we expect to observe from the Earth. 
The predictions in \S~\ref{sec:results} are the distribution of ISOs in a smooth Galaxy-wide distribution, evaluated at the location of the Solar System. 
Gravitational focussing by the Sun will increase the density of ISOs in the inner solar system in a velocity-dependent manner \citep[e.g.][]{Engelhardt_2017,Forbes_2019,Dehnen_2022}, and therefore also in a composition-dependent manner. 
This is due to the fact that we expect compositionally and dynamically distinct populations of ISOs to come from the chemo-dynamically distinct stellar populations in the Milky Way thin and thick disks \citep[c.f.][]{Eubanks_2021}. 
This could handily be modelled when incorporating ISO velocities into the predictions of this framework.

Additionally, the Pan-STARRS near-Earth object survey \citep{Chambers_2016} that detected 1I/\okina Oumuamua, the observations of amateur astronomers such as Gennadiy Borisov who discovered 2I/Borisov, and the LSST which will discover tens more ISOs: all have highly non-trivial selection functions. 
Since in order to be discovered an ISO needs to be detected in multiple observations which can be linked as the same object \citep[e.g.][]{Meech_2017, Schwamb_2023}, these selection functions are dependent on ISO size, approach velocity, perihelion, and composition. 
Composition has a direct link to the detectability of ISOs, as 2I-sized and cometary ISOs will be more likely to be detected, as they will form a coma as they approach the Sun. 
These selection effects will need to be accurately accounted for, in order for the Bayesian framework to produce accurate inferences about the processes affecting the Galactic ISO population.

\section{Conclusion}\label{sec:conc}
In advance of the Vera C. Rubin Observatory Legacy Survey of Space and Time (LSST), we lay out a framework to predict the Galactic distribution of ISOs, using the stellar population of the Milky Way.
Using the method of \cite{Bovy_2016b}, we fit simple density models to a sample of red giants in APOGEE binned in \(\FeH\) and \(\alphaFe\), and use these to evaluate the \textit{sine morte} metallicity distribution of stars throughout the Galaxy's integrated history, across the Galactic disk. 
Under the assumption that the spatial distribution of a population of ISOs will be the same as the \textit{sine morte} distribution of stars which formed them, we use the protoplanetary disk model of \cite{Bitsch_2020} to map the metallicity distribution of stars to the distribution of ISO water mass fractions. 
Localising our model to the Solar neighbourhood, we predict that 95\% of ISOs around the Sun have water mass fraction \(\fHHO\) between 0.07 and 0.51, with a peak around 0.35. 

By considering the distribution of ISOs over the Galactic disk, we show that the well-studied Milky Way metallicity gradient has an equivalent gradient in ISO composition, with the median ISO water mass fraction increasing with distance from the Galactic Centre as the median stellar metallicity decreases. 
This causes the ISO water mass fraction distribution averaged over the Milky Way disk to be wider that the distribution around the Sun. 
Since we also predict higher-metallicity stars produce more ISOs than lower-metallicity stars, the Milky Way metallicity gradient implies that the radial ISO density profile is steeper than the exponential radial stellar density profile, making ISOs much more common in the inner Galactic disk than in the outer disk. 

We also set out a Bayesian inference framework, which can compare predictions of the ISO distribution to the sample observed by the LSST, in order to make inferences about the many astrophysical processes which influence the ISO population. 
To demonstrate its use, we use the composition measurement of 2I/Borisov to calculate a maximum \textit{a posteriori} estimate for the power law slope of the ISO production metallicity dependence, \(\beta\), to be 1.33, with a symmetric 90\% confidence interval of  \((-1.3, 7.2)\).

Encoded in the population of ISOs we observe is a wealth of information about processes through Galactic history, from the evolution of the Milky Way to planet formation. 
The framework set out in this work is a novel approach that will allow us to appreciate how these treasures can further our understanding on both planetary and Galactic scales.

\begin{acknowledgments}
MJH acknowledges support from the Science and Technology Facilities Council through grant
ST/W507726/1. MJH also thanks the LSSTC Data Science Fellowship Program, which is funded by LSSTC, NSF Cybertraining Grant \#1829740, the Brinson Foundation, and the Moore Foundation; his participation in the program has benefitted from this work.
MTB appreciates support by the Rutherford Discovery Fellowships from New Zealand Government funding, administered by the Royal Society Te Ap\={a}rangi.

Funding for the Sloan Digital Sky 
Survey IV has been provided by the 
Alfred P. Sloan Foundation, the U.S. 
Department of Energy Office of 
Science, and the Participating 
Institutions. 
SDSS-IV acknowledges support and 
resources from the Center for High 
Performance Computing  at the 
University of Utah. The SDSS 
website is www.sdss4.org.
SDSS-IV is managed by the 
Astrophysical Research Consortium 
for the Participating Institutions 
of the SDSS Collaboration including 
the Brazilian Participation Group, 
the Carnegie Institution for Science, 
Carnegie Mellon University, Center for 
Astrophysics | Harvard \& 
Smithsonian, the Chilean Participation 
Group, the French Participation Group, 
Instituto de Astrof\'isica de 
Canarias, The Johns Hopkins 
University, Kavli Institute for the 
Physics and Mathematics of the 
Universe (IPMU) / University of 
Tokyo, the Korean Participation Group, 
Lawrence Berkeley National Laboratory, 
Leibniz Institut f\"ur Astrophysik 
Potsdam (AIP),  Max-Planck-Institut 
f\"ur Astronomie (MPIA Heidelberg), 
Max-Planck-Institut f\"ur 
Astrophysik (MPA Garching), 
Max-Planck-Institut f\"ur 
Extraterrestrische Physik (MPE), 
National Astronomical Observatories of 
China, New Mexico State University, 
New York University, University of 
Notre Dame, Observat\'ario 
Nacional / MCTI, The Ohio State 
University, Pennsylvania State 
University, Shanghai 
Astronomical Observatory, United 
Kingdom Participation Group, 
Universidad Nacional Aut\'onoma 
de M\'exico, University of Arizona, 
University of Colorado Boulder, 
University of Oxford, University of 
Portsmouth, University of Utah, 
University of Virginia, University 
of Washington, University of 
Wisconsin, Vanderbilt University, 
and Yale University.
\end{acknowledgments}

\software{NumPy \citep{Harris_2020}; SciPy \citep{Virtanen_2020}; Astropy \citep{AstropyCollaboration_2013, AstropyCollaboration_2018, AstropyCollaboration_2022}; Matplotlib \citep{Hunter_2007}}

\bibliography{export-bibtex}{}
\bibliographystyle{aasjournal}

\end{document}